\documentclass[pre,twocolumn,letterpaper,nofootinbib]{revtex4-2}
\usepackage[utf8]{inputenc}
\usepackage[english]{babel}
\usepackage{url}
\usepackage{amsmath}
\usepackage{graphics}
\usepackage{xcolor}
\definecolor{link}{rgb}{0 0 1}
\usepackage[%
  final,%
  plainpages=false,  
  breaklinks=true,   
  pagebackref=false, 
  hyperindex=false,  
  colorlinks=true,   
  linkcolor=link,    %
  citecolor=link,    %
  menucolor=link,    %
  urlcolor=link,     %
  bookmarks=true,    
  bookmarksnumbered=true,  
  hyperfootnotes=false, 
  pdfcreator={pdfLaTeX},
  pdfstartpage=1]{hyperref}

\def\Vsph{V_\text{sph}}
\def\Vam{V_\text{am}}
\def\Vdeg{V_\text{deg}}
\def\Vtot{V_\text{tot}}
\def\Sams{S_{\text{am,sph}}}
\def\Samdeg{S_{\text{am,deg}}}

\def\dotR{{\dot R}}

\def\fix{{\footref{fn:param}}}
\def\fit{{\phantom{\fix}}}

\def\celsius{\textrm{C}}
\def\micrometer{\mu\textrm{m}}

\def\hour{\textrm{h}}

\def\lenunit{\mathcal{L}}
\def\timeunit{\mathcal{T}}
\begin{document}
\title{Theory for enzymatic degradation of semi-crystalline polymer particles}
\author{Michael Schindler}
\email{michael.schindler@espci.fr}
\affiliation{Gulliver, École Supérieure de Physique et Chimie Industrielles, Paris Sciences et Lettres Université, CNRS, Paris 75005, France}
\author{Hernan Garate}
\affiliation{Gulliver, École Supérieure de Physique et Chimie Industrielles, Paris Sciences et Lettres Université, CNRS, Paris 75005, France}
\affiliation{Laboratoire Sciences et Ingénierie de la Matière Molle, École Supérieure de Physique et Chimie Industrielles, Paris Sciences et Lettres Université, Sorbonne Université, CNRS, Paris 75005, France}
\author{Ludwik Leibler}
\affiliation{Gulliver, École Supérieure de Physique et Chimie Industrielles, Paris Sciences et Lettres Université, CNRS, Paris 75005, France}
\date{January 20, 2026}

\stepcounter{footnote}

\begin{abstract}
In enzymatic recycling or biodegradation of semi-crystalline plastic waste,
crystalline spherulites embedded into an amorphous matrix hinder and slow down
depolymerisation. When the enzymatic depolymerisation temperature exceeds the
glass transition temperature, these spherulites tend to grow. The
depolymerisation process is thus controlled by a competition between erosion of
the amorphous matrix from the particle surface and the growth of recalcitrant
spherulites within the particle bulk and at its surface. We present a geometric
model that captures this competition, together with an algorithm to solve the
equations numerically. Our algorithm introduces a new extension of
Voronoi/Delaunay tessellation in space. We extract the parameters for the model
from experimental data on the enzymatic depolymerization by hydrolase LCC-ICCG
of PET bottle flakes and textile waste, in order to make a prediction of the
observed degradation yield as a function of time. Both the final yield and the
degradation kinetics are correctly predicted. Most importantly, the model
clarifies how and to which extent nucleating agents, impurities, additives,
and/or rapid crystal growth present in the waste can undermine pretreatment
efforts aiming to initiate depolymerisation from a material with a low initial
crystallinity.
\end{abstract}
\maketitle
{\sffamily\noindent
This document is the unedited author's version of a submitted manuscript subsequently accepted for publication in 'Macromolecules'.
To access the final published article, see \url{https://doi.org/10.1021/acs.macromol.5c03444}.}

\section{Introduction}

Enzymatic depolymerization of plastics has emerged as a promising route to
enable a circular economy for polymeric materials
\cite{petdepolymerase,efficientpethydrolases,recyclingimpact,upcycling,enzymepower,engineeredPEThydrolases,auclair25,depablo24}.
By operating under mild, aqueous conditions and exhibiting high selectivity,
enzymes offer a more sustainable alternative to conventional chemical recycling
processes~\cite{fourenzymes}. In typical processes, plastic material is
mechanically broken (milled) into particles of the size of a few hundred
micrometers. These particles are then dispersed in water and brought into
contact with an enzyme, which adsorbs onto the particle surface and catalyzes
successive bond-cleavage reactions that gradually erode the particle. As
depolymerization progresses, soluble monomers or oligomers are released into
the medium. Enzymes capable of cleaving the backbones of multiple classes of
semi-crystalline polymers, including polyesters~\cite{computredesign},
polyamides~\cite{nylon6}, and certain polyurethanes~\cite{polyurethane},
establish biocatalysis as a powerful strategy for plastic recycling.

Enzymatic depolymerization requires sufficient polymer chain mobility, making
reaction temperature a critical parameter. Depolymerization is most effective
at temperatures above the polymer's glass transition temperature~($T_g$) and
below its crystallisation temperature~($T_c$). The optimal temperature depends
on the polymer: Poly(ethylene terephthalate)~(PET) is typically depolymerised
around
$65$--$70^\circ\celsius$~\cite{petdepolymerase,efficientpethydrolases,fourenzymes}
whereas aliphatic polyesters with $T_g$~well below room temperature, such as
poly($\epsilon$-caprolactone) and poly($1,4$-butylene adipate), can be
efficiently degraded at milder temperatures ($30$--$40^\circ\celsius$)
\cite{Hedenqvist22,polycaprolactone,fusing,Welzel03}.

A key feature of most commodity or high performance engineering plastics is
their semi-crystalline nature, deliberately controlled during synthesis and
processing~\cite{Strobl2007,Nurazzi2023}, in order to achieve desirable
properties such as mechanical strength and dimensional
stability~\cite{Meijer2007}, gas barrier performance, and chemical
resistance~\cite{Safandowska2022}. Crystallinity, while crucial for
performance, reduces susceptibility to enzymatic depolymerization: amorphous
domains are amenable to enzyme-catalyzed chain scission, whereas crystalline
regions are densely packed and energetically unfavorable for bond cleavage,
making them
recalcitrant~\cite{Welzel03,petdepolymerase,engineeredPEThydrolases,currentPETminireview}.

The crystallinity problem can be mitigated for polymers that crystallise
slowly, such as PET or~PLA, by an additional pretreatment step before milling.
If the plastic particles are melted and rapidly cooled (quenched) before
milling, then they will be mostly amorphous. The cooling conditions (e.g.~melt
temperature, cooling rate or cooling bath temperature) and polymer molecular
parameters will determine the residual degree of crystallinity and crystallite
organisation and morphology. A~material made of shorter chains, which
crystallise faster than longer chains, exhibits a higher degree of
crystallinity after quenching under identical conditions. Additionally, real
plastic waste streams contain an uncontrolled number of nucleating agents, as
well as other additives such as catalysts, plasticisers, dyes, pigments, impact
modifiers, or fillers, which favour nucleation and crystal growth. For such
highly formulated wastes amorphization by fast melt cooling pretreatment is
more challenging and less efficient than for model polymers or less formulated
or contaminated wastes such as PET~bottles~\cite{ourPNAS}.

The pretreatment by melting and milling serves as a starting point for the
enzymatic depolymerization. The reaction temperature is above~$T_g$ and
below~$T_c$, such that any of the mentioned impurities or preexisting
spherulites inside the otherwise quenched amorphous matrix tend to grow, forming
partly crystalline spherulites which grow and
coalesce~\cite{AntwerpenKrevelen72}. Their growth rate depends on how
close the reaction bath temperature is to~$T_g$ and to~$T_c$. The number of
spherulites and thus also their distance is determined by the density of the
nucleating impurities or the density of preexisting spherulites. What follows
is a competition between spherulite growth (partial crystallisation) and degradation of
amorphous matrix in the reaction bath. It is one of the aims of the theory
presented here to predict how the depolymerization kinetics and reaction yield
will depend on the starting crystalline structure.

The role of initial crystallinity and of its evolution in the reaction bath
in plastics degradation has mainly been investigated in terms of the overall
crystallinity
degree~\cite{ThoHunMey22a,ThoHunMey22b,ThoAlmMey23,WeinbergerEtal_PEF_crystdegree17,RonkvistEtal09,SobkowiczEtal22}.
There are only few experimental studies~\cite{WeiEtal19,RonkvistEtal09}, and we
are aware of one model~\cite{Fabre19}, which consider also the morphology of
the materials while degrading.

\begin{figure}
  \centering
  \includegraphics{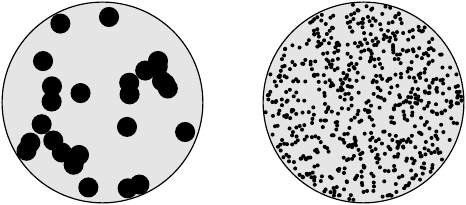}%
  \caption{Sketches representing the spatial configuration of controlled
  preparation at two different temperatures (left:~high, right:~low). They
  serve as starting points for the degradation process. Spherulites in the
  right panel are chosen five times smaller and $5^3$~times more numerous (in
  three dimensions) than in the left panel, but their volume fractions are
  equal.}\label{fig:motivation}
\end{figure}

Beyond the overall crystallinity, the morphology of the spherulites and
their growth dynamics plays a decisive role for the quantitative outcome of
the degradation process. Figure~\ref{fig:motivation} visualises the concept:
The same volume of crystalline material is distributed into different numbers
of spherulites. They have the same crystallinity degree, but they will evolve
differently during depolymerization. Even if the spherulites grow at the same
rate, it is visually evident that the many small ones are closer to each other,
and they will clump together earlier, and thus give less time for the enzymes
to degrade the amorphous matrix between the spherulites. The model presented in
this paper quantifies this difference.

The simultaneous occurrence of depolymerization and crystallization typically
results in a non-monotonic temperature dependence of the yield, with a maximum
at an intermediate temperature \cite{tempdep,compete,ourPNAS}. Due to the
inherent complexity of plastic waste streams, the location of this optimum
cannot be anticipated a priori, emphasizing the need for a theoretical
description that captures the dynamic interplay between crystallization and
enzymatic depolymerization and enables prediction of reaction yields.

Our approach focuses on the initial plastic morphology and explicitly accounts
for thermally activated crystallization during depolymerization. The growing
spherulites block enzymes' access to residual amorphous material and limit the
overall yield. We present a geometric model that quantitatively describes the
competition between amorphous-phase degradation and spherulite growth,
providing both analytical and numerical solutions. The model tracks three key
volumes over time: the spherulitic volume, the degraded amorphous matrix, and
the remaining amorphous volume. Its main outputs are the ultimate fraction of
polymer that will be degraded and the kinetics of this process, quantities that
cannot be inferred from degradation or from crystallization alone. By
combining both effects, the model provides a predictive framework for
understanding and optimizing enzymatic depolymerization under
recycling-relevant conditions. We illustrate its application to~PET from
bottles and textiles~\cite{ourPNAS}, demonstrating how morphological features and processing
history govern reaction yield and kinetics, and suggesting experimental
strategies to validate and exploit the model for process optimization.

Since the model uses only geometrical quantities, such as volumes and volume
fractions, care has to be taken when comparing it with experiments. The growing
spherulites are not entirely crystalline, but consist of stacked crystalline
layers with amorphous polymer chains between them. The internal degree of
crystallinity of a PET~spherulite is typically only~$30$--$40\%$. It is therefore
important to distinguish between the \emph{volume fraction filled by
spherulites} (measured by light-scattering) and the \emph{amount of
crystallinity} (measured by Fourier transform infrared, FTIR, spectroscopy and
by calorimetry). We account for this difference below, in
section~\ref{sec:second_stage_depoly}, where we infer the model parameters from
experimental data.

Here we assume that the enzyme used is sufficiently thermostable so that the
observed temperature dependence reflects polymer mobility and crystallization
phenomena rather than enzyme inactivation. We further assume that plastic
particles are suspended in an aqueous medium and saturated with enzyme, ensuring enzyme availability is not
rate-limiting, and that kinetics are controlled by polymer structure.

\section{A geometric model}
\label{sec:model}
In order to capture the main ingredients of the degradation evolution in the
reaction bath, we formulate a simple Avrami-like~\cite{Avrami39,AvramiReview}
geometric model for it. We take spherical geometry for both the degrading
particle and the growing spherulites. The spherical geometry is not meant to
literally describe spheres in the experiments, the model rather focuses on
their relative overlap. We expect our results to extend also to objects that
have a less well-defined boundary and different shapes. We describe the
degradation of the amorphous matrix of a semi-crystalline particle by a sphere having at time~$t$ the
decreasing radius
\begin{equation}
  \label{eq:Rht}
  R_h(t) = R_{h0} + \dotR_h t, \quad\text{with constant $\dotR_h < 0$.}
\end{equation}
The radius decreasing linearly in time reflects that the volume change of
amorphous matrix material is proportional to the area of surface exposed to
enzymes. Embedded in the amorphous matrix we place spherulites, also modelled
by spheres. The spherulites are assumed to be non-degradable; we discuss this
assumption below, in section~\ref{sec:second_stage_depoly}. Each spherulite has
a given initial radius and is growing, because at the chosen temperature the
surrounding amorphous matrix tends to crystallise and is thus transformed into
spherulite volume. Also here we assume a constant rate, such that a spherulite
radius at time~$t$ is given by
\begin{equation}
  \label{eq:Rst}
  R_s(t) = R_{s0} + \dotR_s t, \quad\text{with constant $\dotR_s > 0$.}
\end{equation}
We take the spherulite positions to be uniformly random distributed within
the amorphous sphere, such that initially they do not stick out. Their number
is denoted by~$N_s$ and the density of centers (number per volume) by~$N_s'$.

\begin{figure}
  \centering
  \includegraphics{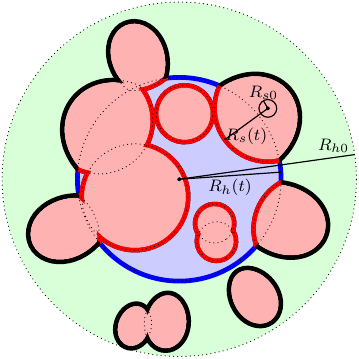}%
  \caption{The three different geometric volumes used in the model, at a
  given time~$t$: Degraded volume~$\Vdeg(t)$ (light green), amorphous
  volume~$\Vam(t)$ (light blue) and spherulite volume~$\Vsph(t)$ (light red).
  The thick solid curves show the interfaces between these volumes.}
  \label{fig:volsurf}
\end{figure}

\begin{figure}
  \centering
  \includegraphics{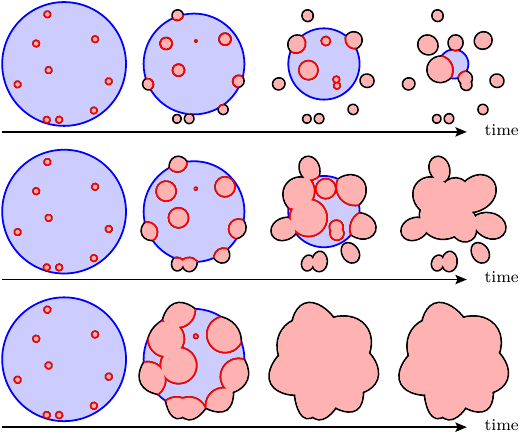}%
  \caption{Evolution in time of the geometry used in the model: The
  amorphous volume (blue) is degraded, the spherulites (red) grow (and possibly
  nucleate) only within the amorphous phase, finally remaining as connected clusters
  of egg-shaped objects (black curves). The three rows correspond to different
  growth rates of spherulites with the same degradation rate of the amorphous phase.}\label{fig:volsurf_evolv}
\end{figure}

As the radii evolve in time, the embedded spherulites come into contact and
grow into each other, and the amorphous matrix is degraded from the outside,
finally leaving clusters of spherulites as a remainder. The resulting geometric
shapes at some intermediate time~$t$ are shown in figures~\ref{fig:volsurf}
and~\ref{fig:volsurf_evolv}. The characteristic egg-like shape of spherulites,
with their long axis oriented towards the center of the particle, is the result
of intersection of a growing and a shrinking sphere.
There are three different volumes in figure~\ref{fig:volsurf}, the spherulite
volume~$\Vsph(t)$ depicted in light~red, the remaining amorphous
volume~$\Vam(t)$ in light~blue, and finally in light~green the
volume~$\Vdeg(t)$ having been degraded up to time~$t$. These three volumes
always add up to the initial volume, which will serve to normalise them,
\begin{equation}
  \label{eq:Vtot}
  \Vam(t) + \Vsph(t) + \Vdeg(t) = \frac{4\pi}{3}R_{h0}^3 =: \Vtot.
\end{equation}
Figure~\ref{fig:volsurf} shows also the interfaces between the different
volumes. The rate at which amorphous volume is converted into spherulite
volume is proportional to the area of the amorphous--spherulite
interface~$\Sams$ (dark~red in figure~\ref{fig:volsurf}),
\begin{equation}
  \label{eq:Shs}
  \frac{d}{dt}\Vsph(t) = \dotR_s\,\Sams(t).
\end{equation}
This quantity is (up to a constant scaling factor corresponding to the internal degree of crystallinity of the spherulites) accessible by calorimetry
experiments, by light-scattering or by Fourier transform infrared (FTIR)
spectroscopy~\cite{FastCrystPET96,CrystPET17,ourPNAS}\footnote{To be more
precise, light-scattering measures the \emph{volume fraction of spherulites}, whereas
Fourier transform infrared (FTIR) spectroscopy measures the total \emph{degree of
crystallinity}. The scaling factor is constant in the temperature range used
here. We thank our reviewer for stressing this point, and we fully account for
the difference between these two quantities below, in
section~\ref{sec:second_stage_depoly}.}. We model the spherulites to be
inert to enzymes: Their part sticking out of the amorphous matrix (thick black
curves in figure~\ref{fig:volsurf}) does not change shape anymore, and the
spherulites grow only inside the amorphous sphere. At the same time they block
access of enzymes to the amorphous matrix and thus effectively reduce the
degradation rate. The degradation of the amorphous volume is thus proportional
to the area of the exposed amorphous--solute interface~$\Samdeg$ (thick
dark~blue curves in figure~\ref{fig:volsurf}),
\begin{equation}
  \label{eq:Shv}
  \frac{d}{dt}\Vdeg(t) = |\dotR_h| \Samdeg(t).
\end{equation}

Equations~\eqref{eq:Rht}--\eqref{eq:Shv} are a system of ordinary differential
equations~(ODE) for the three time-dependent volumes. Before solving them let
us look at two special cases where they are decoupled.

We denote units of length and time by $\lenunit$ and $\timeunit$,
respectively.
\subsection{Degradation alone}
\begin{figure}
  \centering
  \includegraphics{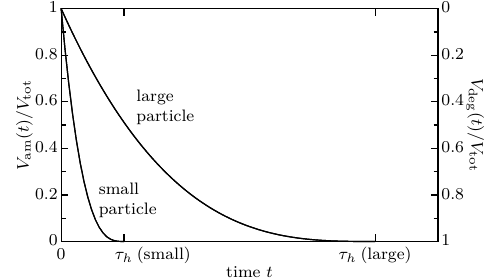}
  \medskip\par
  \begin{tabular}{r|cc|c}
    parameters:\footref{fn:units} & $R_{h0}$     & $\dotR_h$             & $\tau_h$    \\\hline
    small particle                & $\lenunit$   & $-\lenunit/\timeunit$ & $\timeunit$ \\
    big particle                  & $5\lenunit$  & $-\lenunit/\timeunit$ & $5\timeunit$
  \end{tabular}
  \caption{Evolution in time of relative amorphous and of so-far degraded
  volume, without spherulites, according to equations~\eqref{eq:Vtot}
  and~\eqref{eq:amorph_vol}. The timescale of degradation, $\tau_h$~is defined
  in equation~\eqref{eq:tauh}.}\label{fig:evolution_nospherulites}
\end{figure}

In the absence of spherulites, there is only degradation of the wholly amorphous particle, and the relative
volume of the amorphous matrix is simply
\begin{equation}
  \label{eq:amorph_vol}
  \frac{\Vam(t)}{\Vtot} = \left(\frac{R_h(t)}{R_{h0}}\right)^3,
\end{equation}
with the time-dependent radius given by equation~\eqref{eq:Rht}. Small
particles are degraded more quickly, even for the same enzyme activity, because
they expose more surface relative to their volume. This effect can be seen in
figure~\ref{fig:evolution_nospherulites}.

The timescale of degradation is the time on which a purely amorphous particle
vanishes, see equation~\eqref{eq:Rht}. We define it as
\begin{equation}
  \label{eq:tauh}
  \tau_h := \frac{R_{h0}}{|\dotR_h|}.
\end{equation}
%
\subsection{Spherulite growth alone}
\begin{figure}
  \centering
  \includegraphics{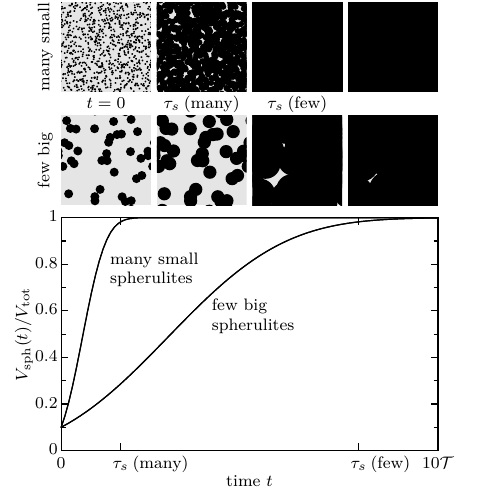}
  \medskip\par
  \begin{tabular}{r|ccc|c}
    parameters:\footref{fn:units} & $R_{s0}$        & $\dotR_s$                 & $N_s'$               & $\tau_s$       \\\hline
    many small                    & $0.02\lenunit$  & $0.03\lenunit/\timeunit$  & $3144.12/\lenunit^3$ & $1.6\timeunit$ \\
    few big                       & $0.10\lenunit$  & $0.03\lenunit/\timeunit$  & $25.15/\lenunit^3$   & $7.9\timeunit$
  \end{tabular}
  \caption{Evolution in time of relative spherulite volume in bulk, according
  to equation~\eqref{eq:avrami_vol}. The timescale of spherulite growth,
  $\tau_s$~is defined in equation~\eqref{eq:taus}. Top panels: spatial sketches
  at chosen times.}\label{fig:evolution_nodegradation}
\end{figure}

The central ingredient of the model places it in the class of \emph{Avrami}
models (sometimes also linked to the name of Kolmogorov and
others)~\cite{AvramiReview}. These models capture the mutual overlap of
spheres, placed randomly in unbounded space at a density~$N_s'$ of centers per
volume. In the limit of many spheres, the common volume of the spheres is found
to be an exponential function of the sum of individual volumes. The volume
fraction $\Vsph(t)/\Vtot$ thus starts at an initial (small) value, then
increases, and finally converges to unity, when all space is filled with
spherulites. The formula of it reads
\begin{equation}
  \label{eq:avrami_vol}
  \frac{\Vsph(t)}{\Vtot}
    = 1 - \exp\Bigl(-N_s'{\textstyle\frac{4\pi}{3}}R_s^3(t)\Bigr).
\end{equation}
%
\footnotetext{\label{fn:units}In the parameters of
figures~\ref{fig:evolution_nospherulites}, \ref{fig:evolution_nodegradation},
and \ref{fig:evolution} we do not fix values for units of length and time yet.
We focus on the difference of timescales, not on their absolute value. We will
fix the units when comparing experimental and simulation data, in
section~\ref{sec:mapping}.}%
This function is plotted in figure~\ref{fig:evolution_nodegradation} for the
two densities of figure~\ref{fig:motivation} (but without boundary and
therefore no degradation). The two different initial states, although having
the same volume fraction and the same growth parameter~$\dotR_s$, converge at
very different timescales.

In order to identify the relevant timescale for spherulite growth, let us first
look at a rather dilute case, where spherulites are mostly well separated
initially ($N_s'\ll1$), at the time when particles were put into the reaction
bath. They all start with the same radius~$R_{s0}$ and grow at the same
rate~$\dotR_s$. After a typical time~$\tau_s$ the spherulites are not well
separated anymore but start to grow into each other. It is
\begin{equation}
  \label{eq:taus}
  \tau_s := \frac{1}{\dotR_s} \biggl(\frac{1.28}{\sqrt[3]{N_s'}} - 2R_{s0}\biggr).
\end{equation}
The quantity in parentheses is referred to as the \emph{ligament thickness} in
the context of toughening of polymers~\cite{CorLei07,Wu88}. For the same
initial spherulite volume fraction $\Vsph(0)/\Vtot \approx N_s'\frac{4\pi}{3}
R_{s0}^3$ we can have different timescales~$\tau_s$, as explained above. For
the examples in figure~\ref{fig:motivation} we expect a ratio of five of their
respective values. If there were no degradation, we would obtain the filling
curves in figure~\ref{fig:evolution_nodegradation}.

When we look at spherulites not in unbounded space but inside the degrading
particle, the Avrami approach leading to equation~\eqref{eq:avrami_vol} becomes
inapplicable for two reasons: First, the spherulite volume inserted into the
exponential function in equation~\eqref{eq:avrami_vol} is to be corrected when
it is only partly inside the particle. Second, and more importantly, there is a
subtle implicit assumption underlying equation~\eqref{eq:avrami_vol}, namely
that the spherulites have to be uniformly distributed in space. The boundary of
the particle breaks this uniformity, and a spherulite close to it will not be
isotropically surrounded by other spherulites. As a result, the
function~\eqref{eq:avrami_vol} does not hold.
\subsection{Coupled growth and degradation}
\label{sec:solving}
In the general case, when spherulites grow within the eroding particle, the
growth curves in figure~\ref{fig:evolution_nodegradation} are limited by the
decreasing curves in figure~\ref{fig:evolution_nospherulites} because
spherulites cannot grow outside the amorphous matrix. At the same time also the
degradation of amorphous matrix is reduced because the spherulites block the
access of enzymes to it.

\begin{figure}
  \centering\vbox{%
  \includegraphics{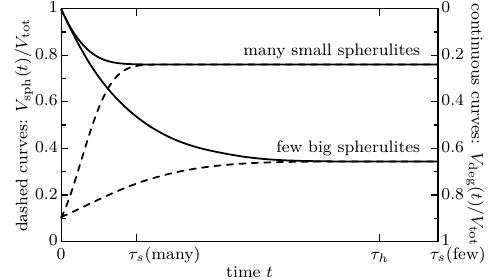}
  \includegraphics{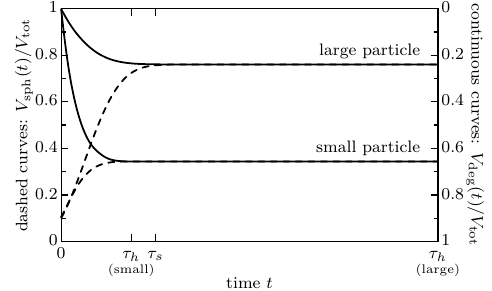}}%
  \medskip\par
  \begin{tabular}{r|ccc|cc}
    parameters:\footref{fn:units} & $R_{s0}$ & $\dotR_s$ & $N_s'$    & $R_{h0}$ & $\dotR_h$ \\\hline
    many small spherul.           & $0.02\lenunit$   & $0.2\lenunit/\timeunit$     & $3145.0/\lenunit^3$  & $ \lenunit$    & $-\lenunit/\timeunit$    \\
    few big spherulites           & $0.10\lenunit$   & $0.2\lenunit/\timeunit$     & $25.2/\lenunit^3$    & $ \lenunit$    & $-\lenunit/\timeunit$    \\
    large particle                & $0.10\lenunit$   & $0.2\lenunit/\timeunit$     & $25.2/\lenunit^3$    & $5\lenunit$    & $-\lenunit/\timeunit$    \\
    small particle                & $0.10\lenunit$   & $0.2\lenunit/\timeunit$     & $25.2/\lenunit^3$    & $ \lenunit$    & $-\lenunit/\timeunit$
  \end{tabular}
  \caption{Numerical solution of the model: evolution in time of the three
  volumes~$\Vsph(t)$, $\Vdeg(t)$, and $\Vam(t)$. Upper panel:~for the
  same~$\tau_h$ and different~$\tau_s$, shown are the two cases of
  figure~\ref{fig:motivation}; $\tau_s$~is varied by varying~$N_s',R_{s0}$,
  while keeping the initial volume constant. Lower panel:~for the same~$\tau_s$
  and different~$\tau_h$, by varying~$R_{h0}$. All volumes are normalised
  by~$\Vtot$. $\Vsph/\Vtot$~is drawn as dashed curves, with values referring to
  the left axis, while $\Vdeg/\Vtot$~is drawn as continuous curves referring to
  the right axis. $\Vam/\Vtot$~is the vertical distance between the pairs of
  curves, according to equation~\eqref{eq:Vtot}.}\label{fig:evolution}
\end{figure}

The model equations~\eqref{eq:Rht}--\eqref{eq:Shv} express this mutual
influence of spherulites and degradation. They result in three volumes as
functions of time, namely in spherulite volume~$\Vsph(t)$, in so-far degraded
amorphous volume~$\Vdeg(t)$, and in remaining amorphous volume~$\Vam(t)$. The
combined dynamics leads to a coupled pair of curves, one for growth, one for
degradation. The two curves join each other at the end of the process when
$\Vam(t)$ vanishes. Figure~\ref{fig:evolution} shows these curves for four
different cases. The three volumes $\Vsph(t)$, $\Vam(t)$, and $\Vdeg(t)$ are
the vertical distances between curves and coordinates axes, respectively. Since
the three volumes add up to the initial total volume~$\Vtot$, plotting them
together in one graph seems the natural way to present the solution, and it
allows also to see their relation at once. At early times, the curves in
figure~\ref{fig:evolution} resemble those in
figures~\ref{fig:evolution_nospherulites} and
\ref{fig:evolution_nodegradation}, but then growth and degradation mutually
influence each other.

Due to the degradation, the final spherulite volume is not the total volume,
but some fraction of it. Its terminal value tells us what portion of the
amorphous matrix has been degraded and what portion crystallised. This fraction
and the typical time when it is reached are the main quantities of interest, in
particular how they depend on the parameters $\dotR_h$, $\dotR_s$, $R_{h0}$,
$R_{s0}$, and~$N_s'$. In reality, these model parameters depend themselves on
the materials used, on the temperature, and on the way the sample was
prepared.

The top panel of figure~\ref{fig:evolution} shows the influence of whether the
initial crystalline volume is split into many small spherulites or few big
ones. We thus come back to the initial question raised around
figure~\ref{fig:motivation}: Is the overall initial crystallinity sufficient to predict
the outcome of the degradation process or are there other important parameters
to consider? Here, the corresponding model parameters $N_s'$~and $R_{s0}$ play
an additional, decisive role. In this simulation we constrain the initial
crystallinity to the same value, but when the same initial volume is split into
many small spherulites, less material is finally degraded. From the different
pairs~$N_s',R_{s0}$ we obtain two different timescales~$\tau_s$ which compare
differently with the given timescale~$\tau_h$.

The lower panel of figure~\ref{fig:evolution} shows the effect of different
timescales~$\tau_h$, as compared to a single~$\tau_s$. This corresponds for
example to different sizes~$R_{h0}$ of particles, obtained from finer/coarser
milling. As expected, small particles are degraded more efficiently, both in
final yield, and in timescale.

In the model we did not specify units of space and time yet. The curves do not
actually depend on the five parameters given, but only on three dimensionless
ones. One choice for the set of dimensionless parameters would be
$\dotR_s/\dotR_h$, $R_{s0}/R_{h0}$, and $N_s'R_{0s}^3$. The same
curves can be obtained also from different sets of parameters: For
example, a bigger particle can be compensated by bigger spherulites, which have
to be less numerous in order to keep the volume fraction unchanged. We will fix
units of length and time in section~\ref{sec:mapping} when extracting model
parameters from experimental data.

\section{Numerical algorithm}
\label{sec:tess}
Our approach to obtain volumes as a function of time is to take a randomly
chosen initial geometric configuration, that is a fixed finite set of sphere
centers and initial radii, and discretise the evolution
equations~\eqref{eq:Rht}, \eqref{eq:Rst} for the radii. We present in this
section a new numerical algorithm devised to efficiently obtain at each
time-step all the required geometric volumes, surface areas, and curve lengths.
In particular, the surface areas on the right-hand sides of
equations~\eqref{eq:Shs} and~\eqref{eq:Shv} are necessary.
\subsection{Double tessellations of space}
The main task of the numerical algorithm is to take into account the overlaps
of spherulites without counting them twice or more. We use a Voronoi
tessellation~\cite{EdeSei86} of space for this, in which every Voronoi cell
corresponds to one sphere. The Voronoi tessellation must be adapted to the
positions and to the radii of the spheres. Our implementation makes use of the
numerical framework CGAL~\cite{cgal:pt-t3-24b,SchMag15} because it allows to
treat \emph{weighted} Delaunay/Voronoi tessellations of spheres which all have
different radii (sometimes also referred to as \emph{regular} tessellation);
this reflects the possibility that spherulites can vary in their initial
radius, or even nucleate at later times.
\begin{figure*}
  \centering
  \includegraphics{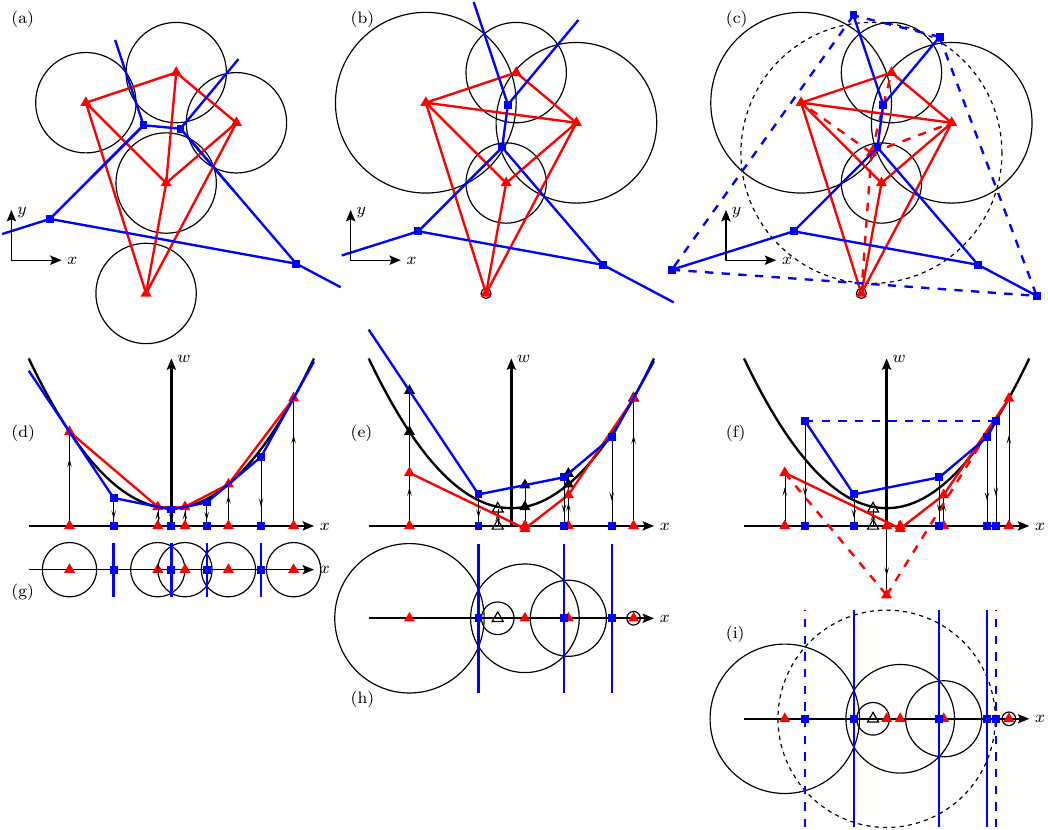}
  \caption{Visualisation of three types of Delaunay/Voronoi tessellation pairs.
  The input spheres are shown as black circles, Delaunay tessellations are
  drawn in red, with little triangles at vertices, and Voronoi tessellations
  are drawn in blue, with little squares at vertices. Input spheres that are
  not part of the Delaunay tessellation have open triangles at their center.
  The left column shows ordinary Delaunay/Voronoi, where all spheres have the
  same radius; the middle column is a \emph{weighted} Delaunay/Voronoi
  tessellation with differently sized spheres; and the right column shows a
  \emph{double} oriented and weighted tessellation with the dashed sphere
  having the others on its inside instead of the outside. The Voronoi cell
  corresponding to the dashed sphere is also drawn with dashed lines. The top
  row gives some examples of two-dimensional tessellations. The middle row
  visualises how Delaunay/Voronoi tessellations can be constructed using
  coordinates lifted to a paraboloid and vertically shifted according to the
  weight of the points (lines with upwards pointing arrows). The Voronoi
  vertices result from a lower convex plane construction in lifted space
  (downwards pointing arrows). The method works in any spatial dimension, but
  visualisation is here limited to one spatial dimension, that is why we show
  the one-dimensional tessellation corresponding to the paraboloid construction
  in the bottom row.}\label{fig:tess}
\end{figure*}%

The weighted tessellation captures all the mutual intersections between growing
spherulites. The usual weighted tessellation, however, only considers spheres
being outside each other, it does not provide the intersections of the
amorphous sphere with the spherulite spheres inside. We present here a new
algorithm, which is an extension of the usual weighted triangulation.
Geometrically, the situation can be seen as a ``double'' oriented and weighted
tessellation, where the Voronoi cells of spherulites are all \emph{contained
in} the Voronoi cell of the amorphous one. This situation is depicted in the
right column of figure~\ref{fig:tess}.

The middle row of figure~\ref{fig:tess} visualises the construction of how a
Delaunay/Voronoi pair of tessellations can be translated into a
lower-convex-hull problem in a space that has one additional ``lifted''
coordinate~\cite{Aur91,EdeSha96}. The visualisation is limited to a single
spatial dimension, plus the lifted one; this one-dimensional tessellation is
shown in the lowest row of figure~\ref{fig:tess}. The extension to two and
three spatial dimensions requires geometric abstraction in the reader's
imagination: for the numerical results in this paper we always used three
spatial dimensions. The left column of figure~\ref{fig:tess} depicts a simple
Delaunay/Voronoi pair of tessellations: The spatial coordinates $(x,y)$ of each
point are lifted to a paraboloid, turning into $(x,y,x^2+y^2)$. The lower
convex hull (depicted in red) of these points defines which point is a Delaunay
vertex, and which ones are neighbors. Bringing this hull down to the $(x,y)$
coordinates defines the Delaunay cells. In each lifted point we construct a
tangent plane to the paraboloid (depicted in blue), which intersect and
together form another convex hull. Bringing this second hull down to the
$(x,y)$ coordinates defines the Voronoi vertices and cells.

The middle column of figure~\ref{fig:tess} shows the effect of different radii
on the construction. We lower the lifted coordinate of input spheres by their
squared radius, also called ``weight'', to $(x,y,x^2+y^2-r^2)$, and take again
the lower convex hull to find the Delaunay cells. The weights have influence on
who is neighbor to whom, and it can even happen that a sphere does not
contribute to the tessellation at all, because it is too small and too close to
a bigger one (see the open triangles in panels e,h). The tangent planes to the
paraboloid are lifted by the sphere weights, thus passing through
$(x,y,x^2+y^2+r^2)$. Again, their intersections define the Voronoi
tessellation. Observe that the lifted planes' intersections coincide with the
intersection lines of spatial circles.

The right column of figure~\ref{fig:tess} shows the addition of the amorphous
sphere (dashed). This is the situation we have to address. Our algorithm
consists in combining two tessellations, one from the spherulites only, and one
into which we insert also the amorphous sphere. We can intersect the two
tessellations only if they have some outermost vertices in common: If
necessary, we surround the whole ensemble with zero-weight dummy points such
that there are always common outermost Delaunay vertices. The two tessellations
form a closed body in lifted coordinates, as is visible in panel~(f) of the
figure. The same is true of the body formed by vertically shifted tangent
planes. Both bodies appear as two overlaying tessellations when projected down
from lifted coordinates to space, see panels~(i) and~(c). As a result, there is
a Voronoi cell dual to the shrinking amorphous sphere (blue dashed), which is
partitioned into Voronoi cells, each dual to one growing spherulite. This
double tessellation allows to measure surface areas of spherulites embedded
into the amorphous matrix, and also the lengths of triple-lines of growing
spherulites, shrinking amorphous particle, and the outer void.
\subsection{Geometric measures}
At every time step, we measure several geometric quantities, namely the volume
of embedded spheres, the area of the amorphous--solute interface, the area of
the amorphous--spherulite interface, and the arc lengths of the
amorphous--spherulite--solute interlines. This is achieved by cutting each
Voronoi cell into wedge-shaped simplices, making use of the right angles
between Delaunay edges and Voronoi faces, and between Delaunay faces and
Voronoi edges. In such an orthogonal simplex the contained spherical volume,
spherical surface, and circular arcs are known
analytically~\cite{MaiLakSas13,SasCorDebSti97,SchMag15}. We add up these
contributions from every simplex of every Voronoi cell of the tessellation. In
the sum, each contribution carries a sign which depends on how the Voronoi face
cuts the Delaunay edge, on how the Delaunay face cuts the Voronoi edge, and on
whether the Voronoi cell corresponds to the amorphous particle or not.

During the time-stepping, the arc lengths of amorphous--spherulite--solute
interlines are integrated to yield the surface area sticking out of the
particle (black in figure~\ref{fig:volsurf}), the amorphous--spherulite
interface area is integrated to give the spherulite volume, and the area of the
amorphous--solute interface is integrated to give the degraded volume.

\section{Finding model parameters from experiments}
\label{sec:mapping}
The most noble aim of any model is to be predictive in the sense that it can
guide experiments. Here, for given experimental parameters the model provides
the final degradation percentage and the timescale. Reversely, if used as a
fitting method, it can provide the model parameters for given experimental
degradation/growth curves. Our plan for this section is to obtain the
degradation parameters $R_{h0}$, $\dotR_h$ from data without spherulite growth,
and to obtain the spherulites parameters $N_s'$, $R_{s0}$, $\dotR_s$ from
growth experiments without degradation. The experimental procedure and
materials are described in reference~\citenum{ourPNAS}.

\subsection{Degradation parameters}
\begin{figure}%
  \centering
  \includegraphics{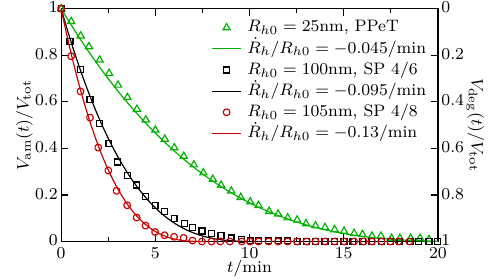}%
  \caption{Degradation of purely amorphous nanoparticles. Experimental data
  taken from Welzel's thesis~\cite[Fig.~5-10]{Welzel03}.
  PPeT~is polypentylene terephthalate, consisting of 1,5-pentanediol and terephthalic acid.
  SP~4/6 is a saturated polyester consisting of 1,4-butanediol and adipic acid.
  SP~4/8 is a saturated polyester consisting of 1,4-butanediol and suberic acid.}%
  \label{fig:welzel510}
\end{figure}

In most cases degradation and growth occur together, which is the whole point
of the model. There is one exception, though. Welzel~\cite{Welzel03} observed
that if the plastic particles are small enough, up to $200\,\textrm{nm}$, they
are totally amorphous. Her two main arguments are that the degradation rates do
not depend on preparation temperatures, and that the cleaving energies are the
same as for liquid polymers~\cite[p.~77]{Welzel03}. For these nanoparticles she
measured the degradation, and we show her data in figure~\ref{fig:welzel510}.
For this degradation, Welzel proposed and confirmed the same degradation
evolution as we do in equation~\eqref{eq:Rht}, where the radius decreases
linearly in time. Data and model match very well.
\subsection{Non-crystalline material in spherulites}
\label{sec:second_stage_depoly}
A major simplification done in the model concerns the definition of spherulite
material. In the model any point in space is either in a spherulite, or in the
amorphous matrix, or degraded, and all material in the spherulites is
effectively taken as ``crystalline''. In experiments this is not absolutely
true~\cite[page~176]{Strobl2007}. When we measure the degree of crystallinity by
isothermal calorimetry, together with a DSC~scan at the end~\cite{ourPNAS},
then even in samples in which spherulites have grown to cover all space, we do not
measure a degree of crystallinity of 100\%, but rather~30\%. This is caused by internal
structure of the spherulites which comprises constrained amorphous polymer
segments~\cite{DiLorenzo24,Wunderlich03} as loops and chains joining crystalline lamellae.
The DSC~experimental method measures only the fusible, crystalline part of their
filling but not the amorphous part.

In order to overcome this discrepancy, we need to rescale the measured
crystallinity to arrive at the spherulite volume. We take the spherulites to be
crystalline to a certain percentage, which does not change with temperature,
and which is constant in time. It can therefore be read off from the terminal
plateau values at~$75^\circ\celsius$ in
figure~\ref{fig:exp_growth_PETparticles}, where we can assume that all space is
covered by spherulites. For example, the PET~bottles treated
at~$75^\circ\celsius$ go up to a terminal crystallinity degree of~$A=27\%$. At
the end, all volume is filled by spherulites, thus~$\Vsph/\Vtot=1$. For those
data lines in figure~\ref{fig:exp_growth_PETparticles} which do not reach the
plateau level in the observation time, we assume the same final level as their
counterparts at higher temperatures: It is the same material and underwent the
same preparation. These values are marked with a symbol (${}\fix$) in
tables~\eqref{tab:exp_growth_PETparticles} and~\eqref{tab:growth_params}.

\begin{figure}%
  \centering
  \includegraphics{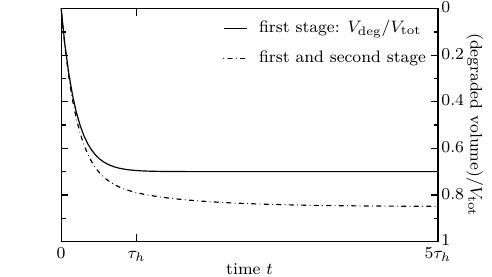}%
  \caption{Schematic sketch of the degradation process happening in two stages:
  First the degradation of main amorphous matrix, then the degradation of
  amorphous material in the spherulites which grew in the first stage and were
  released into the solute. These curves do not come from numerical modeling but are illustrative sketches.}
  \label{fig:fastslow}
\end{figure}
Another consequence of the partial order within spherulites is they are not
completely inert to enzymes but will be degraded as well, however on a longer
timescale. Welzel~\cite{Welzel03} found a timescale at least five times longer
than for the amorphous matrix. We understand the degradation as a two-stage
process in which first the amorphous matrix is degraded and spherulite clusters
are released, and second the spherulite clusters are degraded as well.
Figure~\ref{fig:fastslow} visualises the concept. The second stage means that
in experiments we will not see a well-defined final degradation level, but that
the values continue to change slowly. This can be seen below, in
figure~\ref{fig:exp_cgal_PETparticles_depoly}.
\subsection{Spherulite parameters}
\begin{figure}
  \centering
  \includegraphics{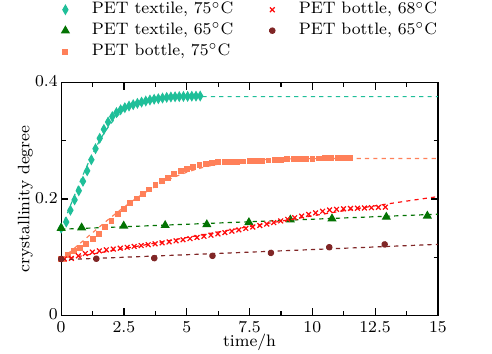}
  \caption{Symbols: Experimental isothermal calorimetry data for crystallinity
  growth in PET particles, in buffer, without degradation. The particles are
  two different plastic waste types, from bottles or from textiles, and are
  crystallised at different temperatures~\cite{ourPNAS}. Dashed curves: Fits
  according to equation~\eqref{eq:fitgrowing3}, resulting in the optimal
  parameters given in
  table~\eqref{tab:exp_growth_PETparticles}.}\label{fig:exp_growth_PETparticles}
\end{figure}
Figure~\ref{fig:exp_growth_PETparticles} shows the experimental curves of
growing crystallinity in PET~particles obtained from bottles and textiles. The
initial particle radius is around $R_{h0}=80\micrometer$. A fit with an Avrami
function needs three parameters,
\begin{equation}
  \label{eq:fitgrowing3}
  \chi(t) = A\Bigl[1-\exp\Bigl(-\frac{4\pi}{3} (B + C t)^3\Bigr)\Bigr],
\end{equation}
with the optimal values\footnote{\label{fn:param}Values marked with a symbol
(\fix) are imposed instead of resulting from a parameter fit.}
\begin{equation}
  \label{tab:exp_growth_PETparticles}
  \begin{tabular}{r|ccc}
    fit of function~\eqref{eq:fitgrowing3}:  & $A$         & $B$         & $C$ \\\hline
    PET textile, $75^\circ\celsius$          & $0.37\fit$ & $0.49\fit$ & $1.5{\times}10^{-1}/\hour$ \\
    PET textile, $65^\circ\celsius$          & $0.37\fix$ & $0.49\fix$ & $2.5{\times}10^{-3}/\hour$ \\
    PET bottle,  $75^\circ\celsius$          & $0.27\fit$ & $0.47\fit$ & $7.2{\times}10^{-1}/\hour$ \\
    PET bottle,  $68^\circ\celsius$          & $0.27\fix$ & $0.47\fix$ & $1.5{\times}10^{-2}/\hour$ \\
    PET bottle,  $65^\circ\celsius$          & $0.27\fix$ & $0.47\fix$ & $3.5{\times}10^{-3}/\hour$.
  \end{tabular}
\end{equation}
In the data time-series, the final level of crystallinity is visible only for
the highest temperature, namely~$75^\circ\celsius$; the parameter~$A$ is then
part of the fitting procedure. For lower temperatures, the final level is not
included in the data, but we assume that it does not depend on the processing
temperature, only on preparation, therefore we impose it in the fitting
procedure. The same applies to the parameter~$B$, which reflects the initial
size of spherulites. In table~\eqref{tab:exp_growth_PETparticles} and in the
following tables, imposed values are marked with a symbol~(\fix).

In order to obtain the model parameters from these fits, we further need a
value for~$N_s'$. Jabarin's publication~\cite{Jabarin82} about the haze in
dried PET~sheets without initial crystallinity gives a distance between
nucleation sites of roughly~$4\micrometer$, for temperatures of
$115^\circ\celsius$ to $130^\circ\celsius$, see Fig.~6 there. Lacking more
precise information, we use this value for~$N_s'$ for the PET~particles.

The correspondence to model parameters is $B=(N_s')^{1/3}R_{s0}$,
$C=(N_s')^{1/3}\dotR_s$, and $A$~translates into $\Vtot$ with a suitable factor
which will not be used. Together with the fit values of
table~\eqref{tab:exp_growth_PETparticles} we now obtain the following model
parameters for spherulites:
\begin{equation}
  \label{tab:growth_params}
  \begin{tabular}{r|ccc}%
                                      & $(N_s')^{1/3}$            & $R_{s0}$              & $\dotR_s$                             \\\hline
      PET textile, $75^\circ\celsius$ & $0.31/(\micrometer)\fix$ & $1.6\micrometer\fit$ & $4.9{\times}10^{-1}\micrometer/\hour$ \\
      PET textile, $65^\circ\celsius$ & $0.31/(\micrometer)\fix$ & $1.6\micrometer\fix$ & $7.9{\times}10^{-3}\micrometer/\hour$ \\
      PET bottle,  $75^\circ\celsius$ & $0.31/(\micrometer)\fix$ & $1.5\micrometer\fit$ & $2.3{\times}10^{-1}\micrometer/\hour$ \\
      PET bottle,  $68^\circ\celsius$ & $0.31/(\micrometer)\fix$ & $1.5\micrometer\fix$ & $4.8{\times}10^{-2}\micrometer/\hour$ \\
      PET bottle,  $65^\circ\celsius$ & $0.31/(\micrometer)\fix$ & $1.5\micrometer\fix$ & $1.1{\times}10^{-2}\micrometer/\hour$.
  \end{tabular}
\end{equation}
\subsection{Depolymerisation hindered by spherulites}
%
\begin{figure}
  \centering
  \includegraphics{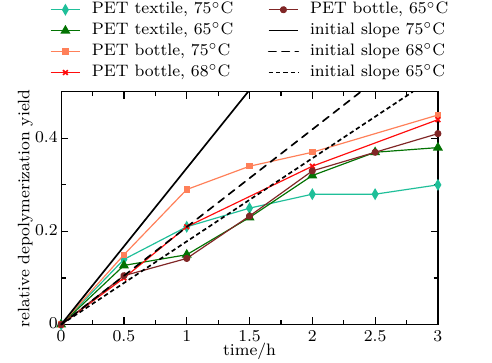}
  \caption{Symbols: Experimental data from depolymerisation of PET particles,
  in buffer~\cite{ourPNAS}. Fits of function~\eqref{eq:fitdepoly2} to the
  initial part of the curves result in the optimal parameters given in
  table~\eqref{tab:exp_depoly_PETparticles}. Their initial slopes, as given in
  table~\eqref{tab:exp_depoly_PETparticles_avg} are visualised by the black
  lines.}\label{fig:exp_depoly_PETparticles}
\end{figure}
The experimental depolymerisation yield from PET~particles is shown in
figure~\ref{fig:exp_depoly_PETparticles}, from the same materials and at the
same temperatures as figure~\ref{fig:exp_growth_PETparticles}.
The depolymerisation was done using the enzyme hydrolase
LCC-ICCG~\cite{petdepolymerase,ourPNAS}. The model's
parameter $\dotR_h$ accounting for depolymerisation is encoded in the initial
slope of the curves in figure~\ref{fig:exp_depoly_PETparticles}. If the initial crystallinity is negligible, then the
initial slope of the depolymerisation curves is simply given by the parameter
combination $\dotR_h/R_{h0}$, see figure~\ref{fig:evolution_nospherulites} for
an example.

If, however, a considerable part of the amorphous--solute interface is covered
by spherulites, they block the enzymes' access to amorphous material. Only a
fraction $\exp\bigl(-\frac{4\pi}{3}N_s'R_{s0}^3\bigr)$ of the shrinking
sphere's surface is exposed to enzymes, and the degradation is reduced by the
same factor. The factor has the same exponential form as seen in
equation~\eqref{eq:avrami_vol}, which is characteristic of an Avrami
argument. Indeed, it has the same origin, only that here we do not have spheres
overlapping in space, but circles overlapping on the particle's surface. The
value of the factor can be calculated from the values given in
table~\eqref{tab:growth_params}, and the initial slope of the curves can be
obtained from figure~\ref{fig:exp_depoly_PETparticles}. To obtain these slopes,
we first fit a smooth function locally to the data at short times, namely the
following cubic function,
\begin{equation}
  \label{eq:fitdepoly2}
  D\Bigl[1-(1 - Et)^3\Bigr].
\end{equation}
The fit yields the optimal parameters $D$~and~$E$,
\begin{equation}
  \label{tab:exp_depoly_PETparticles}
  \begin{tabular}{r|ccc}
    fit of function~\eqref{eq:fitdepoly2}: & $D$    & $E$           & $3DE$        \\\hline
      PET textile, $75^\circ\celsius$      & $0.31$ & $0.34 /\hour$ & $0.31/\hour$ \\
      PET textile, $65^\circ\celsius$      & $0.64$ & $0.096/\hour$ & $0.18/\hour$ \\
      PET bottle,  $75^\circ\celsius$      & $0.42$ & $0.29 /\hour$ & $0.36/\hour$ \\
      PET bottle,  $68^\circ\celsius$      & $0.68$ & $0.1  /\hour$ & $0.21/\hour$ \\
      PET bottle,  $65^\circ\celsius$      & $0.86$ & $0.067/\hour$ & $0.17/\hour$ \\
  \end{tabular}
\end{equation}
The slope at time zero is then given by $3DE$, which equals
$(-3\dotR_h/R_{h0})\exp(-\frac{4\pi}{3}N_s'R_{s0}^3)$ in terms of the model
parameters. The fitting procedure was carried out individually for every
experimental curve. However, the enzyme activity is expected to depend only on
the temperature, and not on the material or the preparation. Indeed, the
fitting nicely groups the initial slopes by their processing temperature. We
can thus average for each temperature,
\begin{equation}
  \label{tab:exp_depoly_PETparticles_avg}
  \begin{tabular}{r|c}
    average initial slope: & $\langle 3DE\rangle$ \\\hline
     $75^\circ\celsius$    & $0.34  /\hour$       \\
     $68^\circ\celsius$    & $0.21  /\hour$       \\
     $65^\circ\celsius$    & $0.18  /\hour$.
  \end{tabular}
\end{equation}
Together with the radius $R_{h0}$~of the particles, which we take to be
$80\micrometer$, we obtain the model parameters for depolymerisation,
\begin{equation}
  \label{tab:degrad_params}
  \begin{tabular}{r|cc}
                                      & $R_{h0}$             & $\dotR_h$                  \\\hline
      PET textile, $75^\circ\celsius$ & $80\micrometer\fix$ & $-13.7  \micrometer/\hour$ \\
      PET textile, $65^\circ\celsius$ & $80\micrometer\fix$ & $ -8.1  \micrometer/\hour$ \\
      PET bottle,  $75^\circ\celsius$ & $80\micrometer\fix$ & $-15.0  \micrometer/\hour$ \\
      PET bottle,  $68^\circ\celsius$ & $80\micrometer\fix$ & $ -8.7  \micrometer/\hour$ \\
      PET bottle,  $65^\circ\celsius$ & $80\micrometer\fix$ & $ -7.2  \micrometer/\hour$.
  \end{tabular}
\end{equation}
%

\section{Results and discussion}
\label{sec:results}
\begin{figure}
  \centering\vbox{%
  \includegraphics{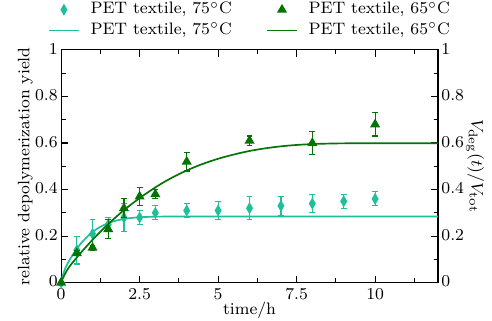}
  \includegraphics{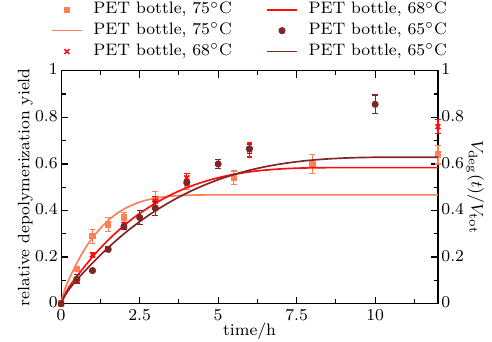}}%
  \caption{Comparison of the numeric model (curves) with experimental
  depolymerisation data (symbols). The symbols refer to the left axis, while
  the curves refer to the right axis. Model parameters are given in
  tables~\eqref{tab:growth_params}
  and~\eqref{tab:degrad_params}.}\label{fig:exp_cgal_PETparticles_depoly}
\end{figure}

In the above section we learned how to get the model parameters for
depolymerisation and for spherulite growth from experimental data. With the
values in table~\eqref{tab:degrad_params} we obtained the parameters $R_{h0}$,
$\dotR_h$ characterising degradation, and with the values in
table~\eqref{tab:growth_params} those characterising spherulites, $N_s'$,
$R_{s0}$, $\dotR_s$. Now we can use both sets of parameters in our model to
make a numerical prediction for combined degradation and spherulite growth.
Figure~\ref{fig:exp_cgal_PETparticles_depoly} shows this prediction for the
depolymerisation of PET~particles from the two different sources of waste and
different temperatures and compares it with the experimental data.

For PET~from textile waste, the model fits the experimental data very nicely,
see the upper panel of figure~\ref{fig:exp_cgal_PETparticles_depoly}. Both the
final yield and the timescale are predicted by the model, at both temperatures.
The match between model and data is a real success of the model and of all its
underlying assumptions. We reiterate here that the Avrami model describes overlap of
spheres with well-defined sharp boundary, which works for spherulites in
textile PET waste, and that in the model the final level of
depolymerisation is only indirectly given by the final level of
crystallisation. Also here the assumption of sharply bounded spherulites
enters. This final level has been accessible from experiments only
for~$75^\circ\celsius$, see figure~\ref{fig:exp_growth_PETparticles}. At
$65^\circ\celsius$ we had to deduce it from the $75^\circ\celsius$ data. And
finally, the rate of depolymerisation has been estimated from the very first
part of the depolymerisation data only. Altogether, the spectacular matching of
model prediction and experimental data in the upper
figure~\ref{fig:exp_cgal_PETparticles_depoly} corroborates the chain of
arguments and approximations we used in the course of sections~\ref{sec:model}
and~\ref{sec:mapping}.

The model slightly underestimates the depolymerisation of PET~textile for times
later than eight hours. Especially the data at~$65^\circ\celsius$ seem to
increase slowly again after having levelled before. We attribute this slow
increase to the second-stage depolymerisation described in
section~\ref{sec:second_stage_depoly}.

For PET~from bottle waste, the comparison of model and experiment is shown in
the lower panel of figure~\ref{fig:exp_cgal_PETparticles_depoly}. Here, the model
matches experimental data up to five hours of depolymerisation. After that,
there are substantial deviations. We think that they have two reasons, first
that the growing spherulites do not have a well-defined boundary and the model
idea of overlapping spheres is thus too simplistic. Second, that these
spherulites have a highly amorphous content, and the second-stage
depolymerisation sets in early and is quite pronounced. We remind the reader of
figure~\ref{fig:exp_growth_PETparticles} where the highest observed
crystallinity degree in PET bottles is only around~$27\%$.

\begin{figure}
  \centering\vbox{%
  \includegraphics{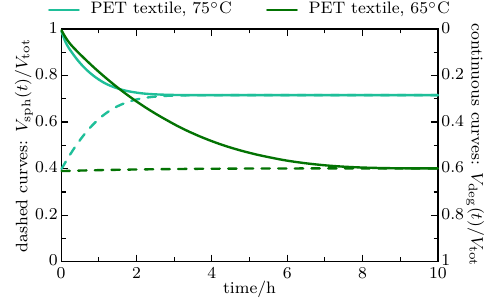}
  \includegraphics{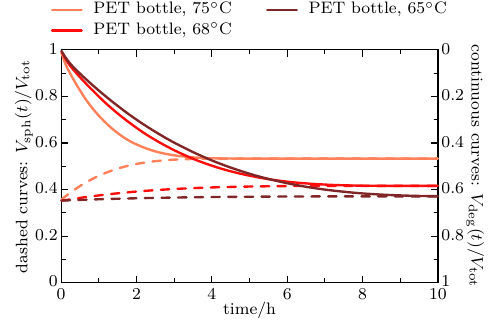}}%
  \caption{Numerical solution of the model, applied to PET~particles: combined
  dynamics of amorphous degradation and spherulite growth. Model parameters are
  given in tables~\eqref{tab:growth_params} and~\eqref{tab:degrad_params}. The
  degradation curves are the same as in
  figure~\ref{fig:exp_cgal_PETparticles_depoly}.}\label{fig:cgal_PETparticles}
\end{figure}

We like to draw the reader's attention to an interesting feature of
figure~\ref{fig:exp_cgal_PETparticles_depoly}: The degradation curves at high
temperature start at a steeper initial slope than those at low temperature, one
would expect also a higher terminal yield. The observed terminal yields,
however, are actually lower---the curves reverse their order. The explanation
for this is the spherulite growth, which reduces the effective
degradation, and which is more effective at higher temperatures. The results in
figure~\ref{fig:exp_cgal_PETparticles_depoly} show that this effect is strong
enough to revert the first intuition from initial degradation slopes. It is
thus useful to show the degradation and the growth curves together, in the same
fashion as we did already in figure~\ref{fig:evolution}. This is done in
figure~\ref{fig:cgal_PETparticles}.

The two timescales $\tau_s, \tau_h$, defined in equations~\eqref{eq:taus},
\eqref{eq:tauh}, provide a useful estimate to describe the curves in
figure~\ref{fig:cgal_PETparticles}. From tables~\eqref{tab:growth_params} and
\eqref{tab:degrad_params} we obtain the following values,
\begin{equation}
  \label{tab:tau_s}
  \begin{tabular}{r|cc}%
                                      & $\tau_s$   & $\tau_h$    \\\hline
      PET textile, $75^\circ\celsius$ & $1.9\hour$ & $5.8 \hour$ \\
      PET textile, $65^\circ\celsius$ & $120\hour$ & $9.9 \hour$ \\
      PET bottle,  $75^\circ\celsius$ & $4.6\hour$ & $5.3 \hour$ \\
      PET bottle,  $68^\circ\celsius$ & $23 \hour$ & $9.2 \hour$ \\
      PET bottle,  $65^\circ\celsius$ & $95 \hour$ & $11.1\hour$.
  \end{tabular}
\end{equation}
When $\tau_s\gg\tau_h$ then spherulite growth is sufficiently slow that nearly
all initially amorphous material will be degraded. The curves
at~$65^\circ\celsius$ in figure~\ref{fig:cgal_PETparticles} are this case. At
$75^\circ\celsius$ there is no clear separation of timescales, and degradation
and growth both contribute to the final outcome. This has been clearly visible
in figure~\ref{fig:evolution}, and we show it again in
figure~\ref{fig:cgal_PETparticles} for the parameters from the PET~degradation
experiments.
\subsection{Varying the initial crystallinity}
In this section we vary the parameters of
tables~\eqref{tab:growth_params} and \eqref{tab:degrad_params} to make
predictions for the depolymerisation yield. We vary those parameters
for which we did not do experiments.
\begin{figure}
  \centering
  \includegraphics{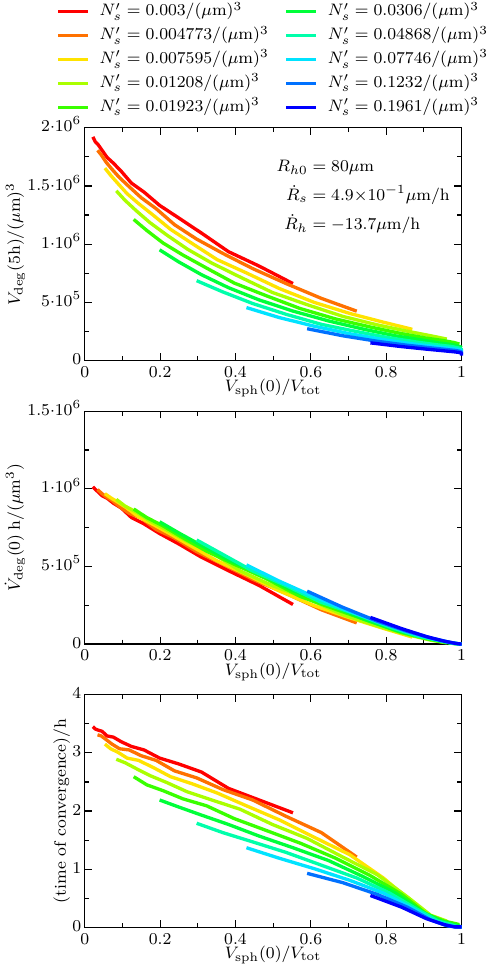}
  \caption{The effect of initial morphology and of initial spherulite volume:
  Numerical solution of the model, applied to PET~particles from textiles,
  treated at~$75^\circ\celsius$. Shown are three chosen quantifiers for the
  yield of monomers, as a function of two parameters $R_{s0}$~and $N_s'$. All
  other parameters are constant as given in tables~\eqref{tab:growth_params}
  and \eqref{tab:degrad_params}. $R_{s0}$~and $N_s'$ are related to the initial
  filling by spherulites~$\Vsph(0)/\Vtot$ via
  equation~\eqref{eq:avrami_vol}.}\label{fig:scanNsprimeRs0}
\end{figure}%
It is common practice to characterise the output of enzymatic degradation by
key quantifiers, such as the yield of monomers after a fixed amount of time, or
the initial yield rate~\cite{Seo25}. These quantifiers are simpler than
the full degradation curve. Figure~\ref{fig:scanNsprimeRs0} shows predictions
from the numerical model for three such quantifiers: for the amount of degraded
monomers after five hours (top panel), for the initial slope of the degradation
curve (middle panel), and for how long the degradation process takes (bottom
panel). For the latter convergence time we take the time when the growth and
degradation curves, such as shown in figure~\ref{fig:cgal_PETparticles},
approach each other closer than $5\%$ for the first time. We base the model
parameters for figure~\ref{fig:scanNsprimeRs0} on those for PET~textile at
$75^\circ\celsius$, as given in tables~\eqref{tab:growth_params} and
\eqref{tab:degrad_params} and then vary both $N_s'$ and $R_{s0}$,
independently. These two parameters determine the initial
filling by spherulites~$\Vsph(0)/\Vtot$ by equation~\eqref{eq:avrami_vol}.

In the top panel of figure~\ref{fig:scanNsprimeRs0} we find again the same
message as in figure~\ref{fig:evolution}: Many small spherulites lead to lower
degradation yield than few big spherulites. Notice that for a given
$\Vsph(0)/\Vtot$, increasing~$R_{s0}$ means decreasing~$N_s'$, and vice-versa.
In the bottom panel of figure~\ref{fig:scanNsprimeRs0} we see that many small
spherulites reach the final levels earlier than few big spherulites. We found
the same message from the conceptual parameters in figure~\ref{fig:evolution}.

The middle panel of figure~\ref{fig:scanNsprimeRs0} quantifies the initial rate
of degradation as a function of the parameters $N_s'$ and~$R_{s0}$. The spread
of these values is much less than in the top panel of the figure. Moreover, the
order of the curves is inverted, that is more and smaller spherulites lead to a
smaller initial slope, despite an increased yield after five hours. The same
inversion is visible in figure~\ref{fig:exp_cgal_PETparticles_depoly}. For the
materials we used, the initial rate of degradation is thus no indicator for the
final yield of degradation.
\subsection{Varying the particle size}
%
\begin{figure}
  \centering
  \includegraphics{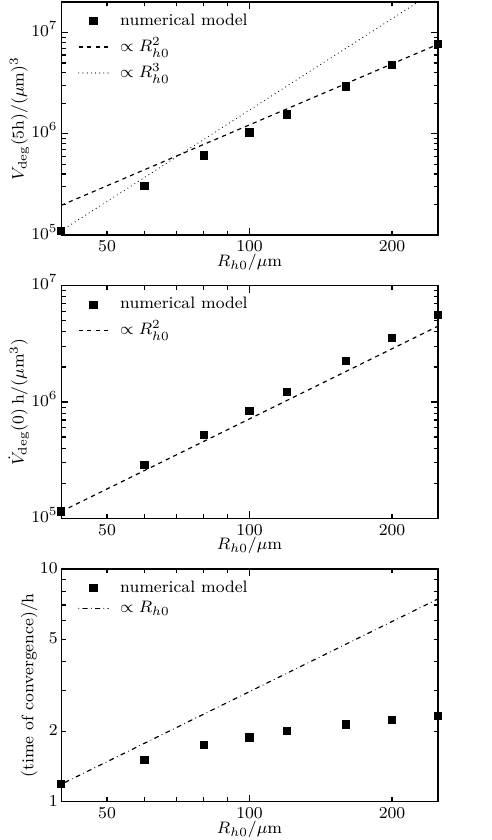}
  \caption{The effect of particle size: Numerical solution of the model,
  applied to PET~particles from textiles, treated at~$75^\circ\celsius$. Shown
  are three chosen quantifiers for the yield of monomers, as a function of the
  parameter~$R_{h0}$. All other parameters are constant as given in
  tables~\eqref{tab:growth_params} and
  \eqref{tab:degrad_params}.}\label{fig:sizedependence}
\end{figure}
\begin{figure}
  \centering
  \includegraphics{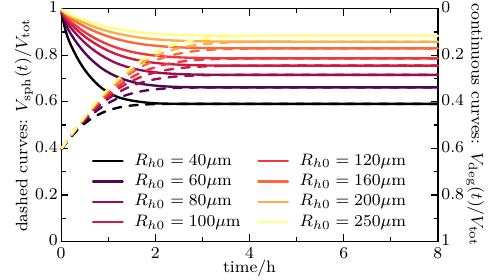}
  \caption{Time-dependent solutions of the model, same as in
  figure~\ref{fig:sizedependence}.}\label{fig:size_evolution}
\end{figure}
Figure~\ref{fig:sizedependence} shows again the three quantifiers, now
varying the particle size~$R_{h0}$. This is an important parameter for
practical application, because it reflects how finely the plastic waste has
been milled. In the top panel of figure~\ref{fig:sizedependence} we see a
crossover of yield from volume-like to surface-like scaling. In larger
particles, the spherulites deep inside the particle have more time to grow, and
if the particle is large enough, they make contact and form a compact object
resisting degradation. In the extreme case only an outer layer of the particle
is degraded, which we see as the quadratic scaling~${\propto}R_{h0}^2$ in the
figure. In smaller particles the spherulites have not enough time to fill space
densely, and the amorphous matrix between them is (partly) degraded. We then
see a volume-like scaling~${\propto}R_{h0}^3$.

The bottom panel of figure~\ref{fig:sizedependence} depicts how long the
degradation process takes, defined as the time when the degradation and the
growth curves approach each other closer than~$5\%$ for the first time. For
large particles, this time seems to converge towards a constant value,
around~$2.3\hour$. This is in accordance with the idea that only an outer layer
of the particle is degraded, leaving a compact object of spherulites. The
thickness of the layer is then around~$32\micrometer$, which explains why at
the left side of the plot the data cross over to linear scaling, in which the
particle is (partly) degraded down to its center.

Instead of showing the absolute degraded volume, it might be more interesting
to see what percentage of a given material one might obtain, depending on how
finely it is milled. In principle, this information is read off the top panel
of figure~\ref{fig:sizedependence}, when comparing the data points with the
line $\propto R_{h0}^3$. Figure~\ref{fig:size_evolution} shows this information
more conveniently, by rescaling the evolving volumes by the total initial
volume of the particle. The data points in figure~\ref{fig:sizedependence} were
obtained from the curves in figure~\ref{fig:size_evolution}. Clearly, small
particles are degraded more quickly, and to a higher fraction.

\section{Summary}
Plastic particles from waste sources are usually heterogeneous, comprising an
amorphous matrix which is degradable by enzymes, and recalcitrant spherulites
embedded in the matrix. Because of the thermal treatment before exposing the
particles to enzymes, the heterogeneity evolves in time. We present a model for
these combined dynamics, which accounts for both the degradation of the
amorphous matrix and the growth of embedded spherulites.

The core of the model is that we treat the degrading particle and the
spherulites as spheres whose radii change in time. We thereby convert the combined
growth/degradation problem into a geometrical problem. The spherulites can
overlap, and they can stick out of the degrading particle. In the latter case
the spherulites block enzymes from accessing the amorphous matrix and thus
influence the rate at which amorphous material is degraded. The overlap of
spherulites places our model in the class of Avrami models. Here, the
spherulite growth is further limited by the shrinking amorphous particle. There
is thus a mutual influence of spherulite growth and degradation of amorphous
matrix.

We thus cannot use the simple exponential Avrami function~\eqref{eq:avrami_vol} to
describe the spherulite volume as a function of time. The finite size of the
degrading particle breaks the assumptions underlying the derivation and the
validity of the simple exponential Avrami function. Instead of trying to derive
an alternative Avrami function, we solve the model
equations~\eqref{eq:Rht}--\eqref{eq:Shv} which operate on an explicit finite set
of spherulites at given positions.

In section~\ref{sec:tess} we present a new numerical algorithm to solve the
model equations and to calculate all required geometrical quantities. The core
of the algorithm is to extend a Voronoi tessellation to a double tessellation,
which covers the degrading particle twice.

In section~\ref{sec:mapping} we show how to obtain the model parameters from
experimental data. This is done for two different materials, each degraded at
different temperatures. We find that the mapping between model and experiment
works very well for particles obtained from PET~textile waste, and it works to
an acceptable degree for those taken from PET~bottle waste.

In section~\ref{sec:results} we use the model to make predictions for the
result of degradation for parameters that we did not vary in the experiments.
In particular, the initial crystallinity (more precisely the volume fraction of spherulites) has an important influence on the
final yield of monomers obtained.

We further use our model to render evident that the initial crystallinity and volume fraction of spherulites is
quite an ambiguous quantity and in general does not allow to predict the
outcome of degradation processes. The same overall volume can
be split into many small or few big spherulites, and their number impacts their
distances from each other, and thus together with their growth rate also the
time when they will form a compact recalcitrant object. We elucidate this idea
first on conceptual parameters in figure~\ref{fig:evolution} and then again in
figure~\ref{fig:scanNsprimeRs0} with the parameters taken from experiments.

From the numerical solution of the model we extract selected quantifiers,
namely the obtained yield after some given time (e.g. five hours), the initial
rate of yield, and the time when the process finishes. We find that the initial
rate of degradation is actually no indicator for the final yield of
degradation as it depends on both the size of the particle and on the size and volume fraction of the internal spherulites.

We further use our model to make a prediction for the influence of the particle
size on the yield. We find two regimes, in the first small particles are
degraded to their center, and in the second large particles have only an outer
shell degraded. We can quantify the crossover between these two regimes, as
well as the thickness of the outer layer.

\bibliography{avrami}

\begin{acknowledgments}
The authors thank Costantino Creton, Andrew Griffiths, Yannick Rondelez,
Matthieu Labousse, and Brian Mansaku for valuable discussions.
\end{acknowledgments}

\end{document}